\documentclass[twocolumn,showpacs]{revtex4}

\usepackage{graphicx}%
\usepackage{dcolumn}
\usepackage{amsmath}
\newcommand{\be}{\begin{equation}}
\newcommand{\ee}{\end{equation}}
\newcommand{\bey}{\begin{eqnarray}}
\newcommand{\eey}{\end{eqnarray}}
\newcommand{\bw}{\begin{widetext}}
\newcommand{\ew}{\end{widetext}}
\newcommand{\ww}{\widetilde}
\newcommand{\ov}{\overline}
\newcommand{\ra}{\rangle}
\newcommand{\la}{\langle}
\newcommand{\br}{ {\bf r} }
\newcommand{\bp}{ {\bf p} }

\begin{document}

 \title {
 Sensitivity of Quantum Motion to Perturbation in Triangle Map
 }

 \author{Wen-ge Wang}

\affiliation{
 Department of Modern Physics, University of Science and Technology of China,
 Hefei 230026, China
 \\ Department of Physics, National University of Singapore, 117542, Singapore
 }

 \date{\today}

 \begin{abstract}

 We study quantum Loschmidt echo, or fidelity, in the triangle map whose classical
 counterpart has linear instability and weak chaos.
 Numerically, three regimes of fidelity decay have been found with respect to the perturbation
 strength $\epsilon$.
 In the regime of weak perturbation, the fidelity decays as $\exp (-c \epsilon^2 t^{\gamma})$ with
 $\gamma \simeq 1.7$.
 In the regime of strong perturbation, the fidelity is approximately a function of
 $\epsilon t^{2.5}$, which is predicted for the classical fidelity [G.~Casati, {\it et al},
 Phys.~Rev.~Lett.~{\bf 94}, 114101 (2005)], and decays slower than power-law decay for long times.
 In an intermediate regime, the fidelity has approximately an exponential decay $\exp (-c' \epsilon t)$.

 \end{abstract}


\maketitle



 \section{Introduction}

 The stability of quantum motion in dynamical systems,
 measured by quantum Loschmidt echo \cite{Peres84}, has attracted much attention
 in recent years.
 The echo is the overlap of the evolution of the
 same initial state under two Hamiltonians with slight difference in the classical limit,
 $ M(t) = |m(t) |^2 $, where
 \be m(t) = \la \Psi_0|{\rm exp}(iHt/ \hbar ) {\rm exp}(-iH_0t / \hbar) |\Psi_0 \ra  \label{mat} \ee
 is the fidelity amplitude.
 Here $H_0$ and $H$ are the unperturbed and perturbed Hamiltonians, respectively,
 $ H=H_0 + \epsilon H_1 $,  with $\epsilon $ a small quantity and $H_1$ a perturbation.
 This quantity $M(t)$ is called fidelity in the field of
 quantum information \cite{nc-book}.

 Fidelity decay in quantum systems whose classical counterparts have
 strong chaos with exponential instability, has been studied well
 \cite{JP01,JSB01,CLMPV02,JAB02,BC02,CT02,PZ02,WL02,VH03,STB03,WCL04,Vanicek04,WL05,WCLP05,GPSZ06}.
 Related to the perturbation strength, previous investigations show
 the existence of at least three regimes of fidelity decay:
 (i) In the perturbative regime in which the typical transition matrix element is smaller than the
 mean level spacing, the fidelity has a Gaussian decay.
 (ii) Above the perturbative regime, the fidelity has an exponential decay with a rate
 proportional to $\epsilon^2$, usually called the Fermi-golden-rule (FGR) decay of fidelity.
 (iii) Above the FGR regime is the Lyapunov regime in which $M(t)$ has usually an
 approximate exponential decay with a perturbation-independent rate.

 Fidelity decay in regular systems with quasiperiodic motion in the
 classical limit has also attracted much attention
 \cite{PZ02,JAB03,PZ03,SL03,Vanicek04,WH05,Comb05,HBSSR05,GPSZ06,WB06,pre07}.
 For single initial Gaussian wavepacket,  the fidelity has
 been found to have initial Gaussian decay followed by power law decay\cite{PZ02,WH05,pre07}.

 Meanwhile, there exists a class of system which lies between the two classes of system mentioned above,
 namely, between chaotic systems with exponential instability and regular systems
 with quasiperiodic motion.
 One example of this class of system is the triangle map proposed by Casati and Prosen \cite{triangle}.
 The map has linear instability with vanishing Lyapunov exponent,
 but can be ergodic and mixing with power-law decay of correlations.
 The classical Loschmidt echo in the triangle map has been studied recently
 and found behaving differently from that in systems with exponential instability
 and in systems with quasiperiodic motion \cite{c-fid-tri}.
 This suggests that the decaying behavior of fidelity in the quantum triangle map may be
 different from that in the other two classes of system as well.
 In this paper, we present numerical results which confirm this expectation.

 Specifically, like in systems possessing strong chaos,
 in the triangle map three regimes of fidelity decay are found
 with respect to the perturbation strength: weak, intermediate and strong.
 However, in each of the three regimes, the decaying law(s) for the fidelity in the triangle map has
 been found different from that in systems possessing strong chaos.
 In section II, we recall properties of the classical triangle map
 and discuss its quantization.
 Section III is devoted to numerical investigations for the laws of fidelity decay
 in the three regimes of perturbation strength.
 Conclusions are given in section IV.

 \section{Triangle map}

 \begin{figure}
 \includegraphics[width=\columnwidth]{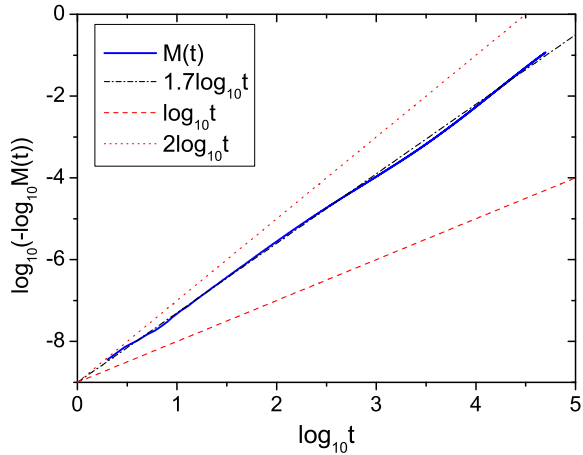}
 \caption{ (color online).
 Averaged fidelity at weak perturbation, $\sigma =10^{-4}$(solid curve),
 with average taken over 50 initial point sources chosen randomly, $N=2^{12}=4096$.
 The dashed-dotted straight line has a slope 1.7, showing that $\log_{10}\ov M(t)$ is approximately
 a function of $t^{1.7}$.
 For comparison, we also show two straight lines (dashed and dotted) with
 slopes 1 and 2, respectively.
 } \label{fig-s0001-t1p7}
 \end{figure}

 On the torus $(r,p) \in {T}^2 = [-\pi ,\pi ) \times [-\pi ,\pi )$,
 the triangle map is
 \bey \nonumber p_{n+1} = p_n + \alpha \ \text{sgn} (r_n)+ \beta , \hspace{1cm} (\text{mod} 2\pi )
 \\ r_{n+1} = r_n + p_{n+1} ,  \hspace{1cm} (\text{mod} 2\pi ) \label{map} \eey
 where $\text{sgn}(r) = \pm 1 $ is the sign of $r$ for $r \ne 0$ and
 $\text{sgn}(r) =0$ for $r=0$ \cite{triangle}.
 Rich behaviors have been found in the map:
 For rational $\alpha /\pi$ and $\beta /\pi$, the system is pseudointegrable.
 With the choice of $\alpha =0$ and irrational $\beta / \pi$, it is ergodic but not mixing.
 Interestingly, for incommensurate irrational values of $\alpha /\pi$ and $\beta /\pi$,
 the dynamics is ergodic and mixing.
 In our numerical calculations, we take $\alpha = \pi^2 $ and $\beta = (\sqrt{5}  -1)\pi /2$,
 {for which $(\beta / \alpha )$ is an irrational number, the golden mean divided by $\pi$,
 and the map is ergodic and mixing.}

 The triangle map (\ref{map}) can be associated with the Hamiltonian
 \bey H = \frac 12 \ww p^2 + V(r) \sum_{n=-\infty }^{\infty } \delta (t-nT), \label{H} \eey
 where $ V(r) = - \ww \alpha |r| - \ww \beta r$ and $T$ is the period of kicking.
 It is easy to verify that the dynamics produced by this Hamiltonian gives the map (\ref{map})
 with the replacement $p=T\ww p, \alpha = T\ww \alpha $, and $\beta =T\ww \beta $.

 \begin{figure}
 \includegraphics[width=\columnwidth]{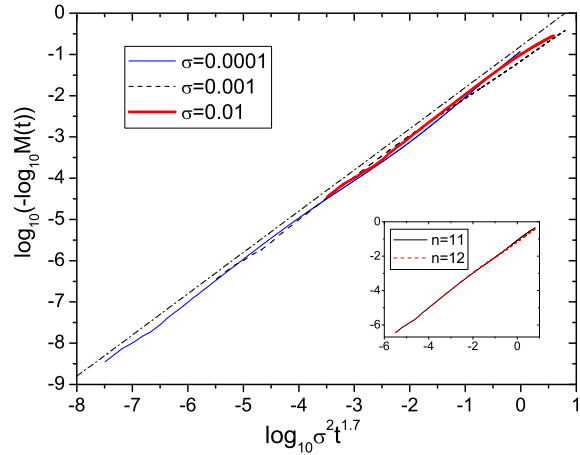}
 \caption{ (color online).
 Averaged fidelity at three weak perturbation strengths, $\sigma =10^{-4}$(thin solid curve), $10^{-3}$
 (dashed curve), and $10^{-2}$(thick solid curve),
 with average taken over 50 initial point sources chosen randomly, $N=2^{12}=4096$.
 The dashed-dotted straight line represents $M_1(t)$ in Eq.~(\ref{ctgamma})
 with $\gamma =1.7$ and $c$ as an adjusting parameter.
 Inset: Fidelity of $\sigma =10^{-3}$ and $N=2^{n}$;
 the two curves are almost indistinguishable.
 } \label{fig-s0001-tlog}
 \end{figure}

 The classical map can be quantized by the method of quantization on torus
 \cite{HB80-q-tori, FMR91,WB94,Haake}.
 Schr\"{o}dinger evolution under the Hamiltonian in Eq.~(\ref{H})
 for one period of time is given by the Floquet operator
 \be \label{U1} U = \exp \left [ -\frac i2 ({\hat{\ww p}})^2T \right ]
 \exp [-i V( {\hat r}) ], \ee
 where we set $\hbar =1$ in Schr\"{o}dinger equation.
 In this quantization scheme, an effective Planck constant $\hbar_{\rm eff}=T$ is introduced.
 It has the following relation to the dimension $N$ of the Hilbert space,
 \be \label{h} N h_{\rm eff} =4\pi^2, \ee
 hence, $\hbar_{\rm eff} = 2\pi / N$.
 In what follows, for brevity, we will omit the subscript eff of $\hbar_{\rm eff}$.
 Eigenstates of $\hat{r} $ and $\hat p$ are discretized,
 $\hat{r}|j\ra = j \hbar |j\ra $ and $\hat{p}|k\ra = k \hbar |k\ra $,
 with  $j,k =-N/2,-N/2+1,\ldots ,0,1, \ldots , (N/2)-1$.
 Then,
 {making use of the above discussed relations among $\ww p, p, T, \ww \alpha , \alpha ,
 \ww \beta , \beta $, in particular, $T=\hbar $},
 the Floquet operator in Eq.~(\ref{U1}) can be written as
 \be \label{U} U = \exp \left [ -\frac {i}{2\hbar} ({\hat{ p}})^2 \right ]
 \exp \left [ \frac{i}{\hbar} (\alpha |\hat r| +\beta \hat r) \right ] . \ee
 In numerical computation, the time evolution
 $ |\psi (t)\ra = U^t |\psi_0\ra $ is calculated by the fast Fourier transform (FFT) method.

 The fidelity in Eq.~(\ref{mat}) involves two slightly different Hamiltonians,
 unperturbed and perturbed.
 In this paper, for an unperturbed system with parameters $\alpha $ and $\beta $,
 the perturbed system is given by
 \be \alpha  \to \alpha  + \epsilon \ \ \ \ \beta  \to \beta   . \ee
 Without the loss of generality, we assume $\epsilon \ge 0$.
 The parameter $\sigma =(\epsilon / \hbar )$ can be used to characterize the strength of quantum
 perturbation.

 \begin{figure}
 \includegraphics[width=\columnwidth]{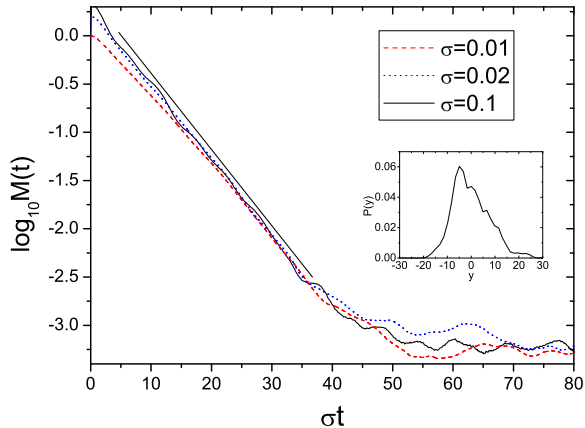}
 \caption{ (color online).
 Variation of the averaged fidelity with $\sigma t$ for  $\sigma =0.01, 0.02$ and 0.1,
 with average taken over 100 initial point sources chosen randomly, $N=4096$.
 The solid straight line is drawn for a comparison with linear dependence on $\sigma t$.
 For $\sigma = 0.02$ and 0.1, $\log_{10} \ov M(t)$ is approximately a linear function of $ \sigma t$,
 before it becomes close to the saturation value.
 Inset: The distribution $P(y)$ for the action difference $\Delta S$ at $t=40$,
 where $y=(\Delta S -\la \Delta S \ra )
 / \epsilon $ and $\la \Delta S \ra$ is the average value of $\Delta S$.
 It is calculated by taking randomly $10^7$ initial points in the phase space.
 $P(y)$ does not have a Gaussian shape.
 } \label{fig-mt-s001-01}
 \end{figure}

 \section{Three regimes of fidelity decay}

 \subsection{Weak perturbation regime}

 Let us first discuss weak perturbation.
 As mentioned in the introduction, in systems with strong chaos in the classical limit,
 the fidelity has a Gaussian decay under sufficiently weak perturbation.
 The Gaussian decay is derived by making use of the first order perturbation theory for eigensolutions
 of $H$ and $H_0$ and the random matrix
 theory for $\Delta E_n \equiv E_n-E^0_n$, where $E_n$ and $E^0_n$ are
 eigenenergies of $H$ and $H_0$, respectively.
 Numerical results in Ref.~\cite{EKW05} show agreement of the spectral
 statistics in the triangle map with the prediction of random matrix theory,
 hence, at first sight, Gaussian decay might be expected for the fidelity decay
 in the weak perturbation regime of the triangle map.

 However, our numerical results show a non-Gaussian decay of fidelity for small perturbation.
 An example is given in Fig.~\ref{fig-s0001-t1p7} for $\sigma=10^{-4}$.
 To obtain relatively smooth curves for fidelity,
 average has been taken over 50 initial point sources (eigenstates of $\hat r$) chosen randomly.
 This figure, plotted with $\log_{10} \left (-\log_{10} \ov M(t)\right )$ versus $ \log_{10} t $,
 shows clearly that $\log_{10}\ov M(t)$ is approximately proportional to $t^{1.7}$ (the
 dashed-dotted straight line), while is far from the Gaussian case of $t^2$ and the
 exponential case of $t$ represented by the dotted and dashed lines, respectively.

 Furthermore, we found that the averaged fidelity $\ov M(t)$ can be fitted well by
 \be \label{ctgamma} M_1(t) = \exp (-c \sigma^2 t^{\gamma }) \ee
 with $\gamma \simeq 1.7$ and $c$ as a fitting parameter.
 In Fig.~\ref{fig-s0001-tlog}, we show fidelity decay for three different values of $\sigma $.
 With the horizontal axis scaling with $\log_{10}\sigma^2 t^{1.7 }$,
 the three curves corresponding to the three values of $\sigma $
 are hardly distinguishable in their overlapping regions (except for long times).
 Note that, to show clearly the dashed-dotted straight line which represents
 $M_1(t)$ in Eq.~(\ref{ctgamma}),
 we have deliberately adjusted a little the best-fitting value of $c$ such that the dashed-dotted line
 is a little above the curves of the fidelity.

 \begin{figure}
 \includegraphics[width=\columnwidth]{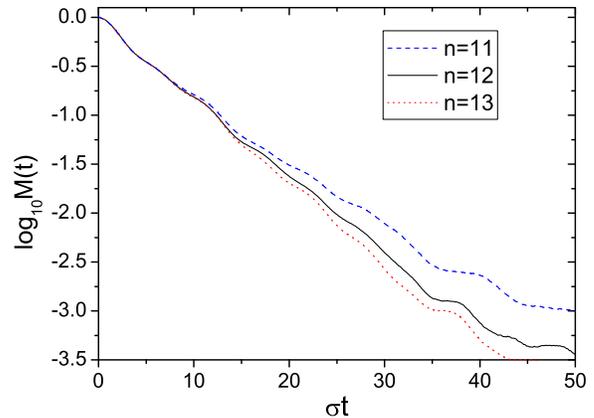}
 \caption{ (color online).
 Fidelity decay for $\sigma =0.1$ and $N=2^{n}$,
 averaged over 100 initial point sources.
 } \label{fig-s0p1-n11-n12}
 \end{figure}

 In the inset of Fig.~\ref{fig-s0001-tlog}, we show curves of fidelity for the same $\sigma$
 but different values of $\epsilon $ and $N$.
 The two curves are very close, supporting the assumption
 that $\epsilon $ and $N$ appear in the form of the
 single variable $\sigma $ as written on the right hand side of Eq.~(\ref{ctgamma}).
 This dependence of $\ov M(t)$ on the variable $\sigma $ for sufficiently small $\sigma $
 can be understood in a first-order perturbation treatment of fidelity,
 as shown in the following arguments.

 Let us consider a Hilbert space with sufficiently large dimension $N$
 and make use of arguments similar to those used in Ref.~\cite{CT02}
 for deriving the Gaussian decay,
 but without assuming the applicability of the random matrix theory.
 It follows that, for times not very long, the averaged fidelity (averaged over initial states)
 is mainly determined by
 $\la \exp (-i\Delta \omega_n t ) \ra$, where $\Delta \omega_n =\omega_{n} -\omega_n^0$
 and $\la \ldots \ra $ indicates average over the quasi-spectrum.
 Here $\omega_n^0$ is an eigen-frequency of the Floquet operator $U$ in Eq.~(\ref{U})
 and $\omega_n$ is the corresponding eigen-frequency of $(U e^{i\sigma |r|})$.
 For large $N$, $\la \exp (-i\Delta \omega_n t ) \ra$
 can be calculated by making use of the distribution of $\Delta \omega_n $.
 Since the two Floquet operators $U$ and $(U e^{i\sigma |r|})$ differ by $e^{i\sigma |r|}$,
 the distribution of $\Delta \omega_n $ is approximately a function of $\sigma $.
 Then, $M(t)$ is approximately a function $\sigma $.

 Finally, we give some remarks on the value of $\gamma $.
 When $\Delta \omega_n$ has a Gaussian distribution, $\ov M(t)$ has a Gaussian decay with $\gamma =2$,
 as in the case of systems possessing strong chaos.
 In the triangle map, the non-Gaussian decay of fidelity discussed above implies
 that $\Delta \omega_n$ does not have a Gaussian distribution.
 Other types of distribution may predict values of $\gamma $ different from 2, in particular,
 a L\'{e}vy distribution would give $\gamma <2$ in agreement with our numerical result.
 We also remark that the results here are not in confliction with numerical results of Ref.~\cite{EKW05},
 in which only the statistics of $\omega_n$ (not that of $\Delta \omega_n$) is found
 in agreement with the prediction of random matrix theory.

 \subsection{Intermediate perturbation strength}

 \begin{figure}
 \includegraphics[width=\columnwidth]{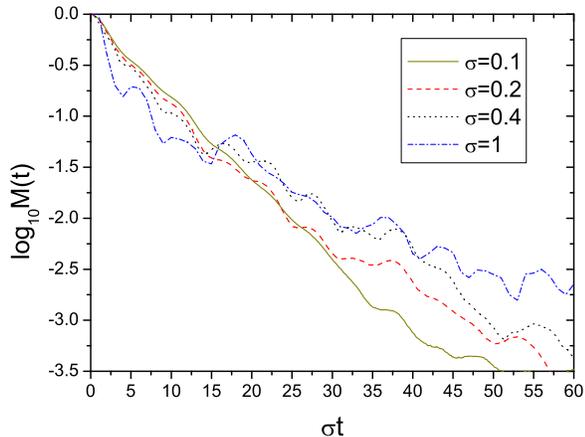}
 \caption{ (color online).
 Averaged fidelity of $\sigma$ from 0.1 to 1, with average taken over 1000 randomly chosen
 initial pointer sources, $N=2^{14}=16384$.
 For $\sigma =0.2$ and above, the averaged fidelity obeys a decaying law which is different from
 that in Eq.~(\ref{Mt-sigma-t}), in particular, it is not a function of $(\sigma t)$.
 } \label{fig-mt-s01-1-poi-st}
 \end{figure}

 With increasing perturbation strength, exponential decay of $\ov M(t)$ appears
 (see Fig.~\ref{fig-mt-s001-01}).
 For $\sigma $ from 0.02 to 0.1, after some initial times and before approaching
 its saturation value, the fidelity decays as
 \be M_2(t) = \exp (- a \sigma t), \label{Mt-sigma-t} \ee
 with $a$ as a fitting parameter.
 Numerically, we found that $a \approx 0.08$.
 The decay rate is proportional to $(\sigma t)$, unlike in the FGR decay found in systems
 with strong chaos,
 \be M_{\rm FGR}(t) \sim \exp (-2 \sigma^2 K_E t), \label{FGR} \ee
 where $K_E$ is the classical action diffusion constant \cite{CT02}.
 The curves of $\sigma =0.02$ and 0.1 in Fig.~\ref{fig-mt-s001-01} are quite close,
 while that of $\sigma =0.01$ has some deviation from the two.
 This implies that the $\exp (- a \sigma t)$ behavior of $\ov M(t)$ appears
 between $\sigma =0.01$ and 0.02.
 Note that vertical shifts have been made for the two curves of
 $\sigma =0.02$ and 0.1  in Fig.~\ref{fig-mt-s001-01} for better comparison.

 \begin{figure}
 \includegraphics[width=\columnwidth]{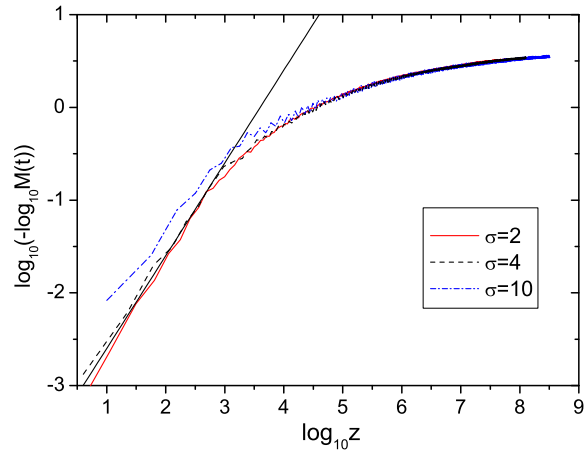}
 \caption{(color online).
 Averaged fidelity at strong perturbation,
 with average taken over 1000 randomly chosen initial Gaussian wavepackets,
 $N=2^{17}=131072$.
 $z=\epsilon t^{2.5}/\hbar $ with $\hbar $ fixed in this figure.
 The solid line represents a curve $\exp (-c \epsilon t^{2.5})$,
 where the fitting parameter $c$ is determined from comparison with the two curves
 of $\sigma =2$ and 4 in the small-$z$ region.
 } \label{fig-lglgm}
 \end{figure}

 The origin of the non-FGR decay of fidelity
 in this regime of perturbation strength, may come from weak chaos.
 In fact, in another system which also possesses weak chaos in the classical limit,
 namely, the sawtooth map in some parameter regime, linear dependence of the decaying rate
 on $\sigma$ has also been observed in the intermediate perturbation regime \cite{WCL04,WL05,foot1}.
 In this regime of perturbation strength, the semiclassical theory predicts that,
 in the first order classical perturbation theory,
 the averaged fidelity is given by \cite{WCL04}
 \bey  \ov M(t) \simeq \left | \int d\Delta S e^{i\Delta S/ \hbar }
 P(\Delta S)\right |^2,   \label{Mp-ps} \eey
 where
 $ \Delta S(\bp_0 , \br_0 ; t) = \epsilon  \int_0^t dt' H_1[(\br (t')]$
 is the action difference of two the classical trajectories starting at the same
 point $(\bp_0 , \br_0)$ in the two systems,
 with $H_1$ evaluated along one of the two trajectories,
 and $P(\Delta S)$ is the distribution of $ \Delta S(\bp_0 , \br_0 ; t)$.
 In systems possessing strong chaos, $P(\Delta S)$ may have a Gaussian form, which implies
 the FGR decay for the fidelity.
 In the triangle map, $P(\Delta S)$ is not a Gaussian distribution
 as shown in the inset of Fig.~\ref{fig-mt-s001-01},
 hence, the fidelity does not have the FGR decay with a rate proportional to $\sigma^2$.

 It is difficult to find an analytical expression for $P(\Delta S)$,
 hence, we can not derive Eq.~(\ref{Mt-sigma-t}) analytically.
 However, a qualitative understanding of the $(\sigma t)$-dependence of $\ov M(t)$ can be
 gained, as shown in the following arguments.
 Equation (\ref{Mp-ps}) shows that the time-dependence of fidelity decay comes mainly from
 the dependence of $P(\Delta S)$ on time.
 In the case of strong chaos, $\Delta S$ behaves like a random walk, hence,
 $P(\Delta S)$ has a Gaussian form with a width increasing as $\sqrt t$ \cite{CT02}.
 Since $\Delta S \propto \epsilon $, the width of $P(\Delta S)$ is a function of $(\epsilon \sqrt t)$;
 then, Eq.~(\ref{Mp-ps}) gives the FGR decay of $\ov M(t)$ which depends on $(\sigma^2t)$.
 In the case of the triangle map, due to the linear instability of the map, it may happen
 that the width of $P(\Delta S)$ increase linearly with $ t$
 in some situations when $t$ is not very long.
 This implies that the width of $P(\Delta S)$ may be a function of the variable $(\epsilon t)$.
 Then, it is possible for $\ov M(t)$ to be approximately a function of $(\sigma t)$.

 Equation (\ref{Mp-ps}) predicts that, up to the first order classical perturbation theory,
 the dependence of $\ov M(t)$ on $\epsilon $ and $\hbar $
 takes the single variable $\sigma =\epsilon / \hbar $.
 Numerically we found that this is approximately correct, as shown in Fig.~\ref{fig-s0p1-n11-n12}.
 Specifically, for fixed $\sigma =0.1$, $\ov M(t)$ of $N=2^{11}$ and of $N=2^{12}$ separate at about $t=15$.
 Indeed, for long times $t$, higher order contributions in the classical perturbation theory may
 need consideration and $\ov M(t)$ may depend on $\epsilon  $ and $\hbar $ in a different way.
 For larger $N$, hence smaller $\hbar $, the agreement becomes better,
 e.g., $\ov M(t)$ of $N=2^{12}$ is closer to $N=2^{13}$ than to $N=2^{11}$.

 When $\sigma $ goes beyond 0.1,
 the exponential decay of $\ov M(t)$ expressed in Eq.~(\ref{Mt-sigma-t}) disappears,
 in particular, the dependence of $\ov M(t)$ on $\sigma $ and $t$ does not take the form of $(\sigma t)$
 (see Fig.~\ref{fig-mt-s01-1-poi-st}).
 Meanwhile fluctuations of $\ov M(t)$ becomes larger and larger with increasing $\sigma $
 for initial point states.
 For example, Fig.~\ref{fig-mt-s01-1-poi-st} shows that $\overline M(t)$ of $\sigma =1$
 has considerable fluctuations even after averaging over 1000 initial point sources.
 Taking initial Gaussian wavepackets, the fluctuations can be much suppressed.

 \begin{figure}
 \includegraphics[width=\columnwidth]{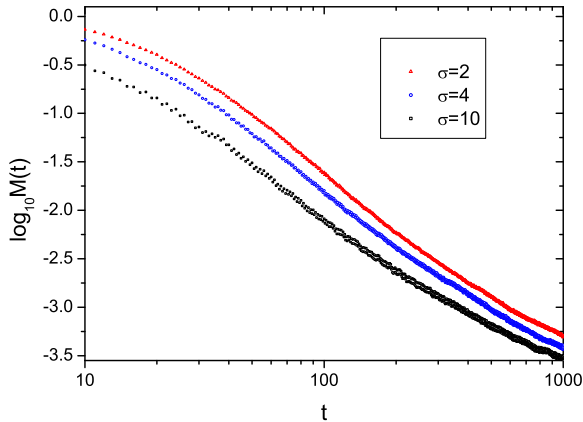}
 \caption{ (color online).
 Averaged Fidelity for strong perturbation, from top to bottom, $\sigma =2,4$ and 10.
 The average is taken over 1000 initial Gaussian wavepackets chosen randomly
 and over time from $t-2$ to $t+2$. $N=2^{17}$.
 The time axis is plotted in the logarithm scale.
 It shows that the long time decay of fidelity is slower than power law decay.
 } \label{fig-mt-sgt1-tavg}
 \end{figure}

 \subsection{Strong perturbation regime}

 The triangle map has vanishing Lyapunov exponent, hence, its fidelity may not have
 the perturbation-independent decay
 which has been observed at strong perturbation in systems possessing exponential
 instability in the classical limit
 \cite{JP01,BC02,STB03,WCLP05}.
 To understand fidelity decay in the triangle map, it is helpful to recall results
 about the classical fidelity given in \cite{c-fid-tri}.
 In the classical triangle map, the classical fidelity decays as
 $M_{cl}(t) \sim \exp (-c \epsilon t^{2.5})$
 for initial times when $M_{cl}(t)$ remains close to one,
 and has an exponential decay $\exp (-c' \epsilon^{2/5}t)$ for longer times.
 The interesting feature is that the classical fidelity depends
 on the same scaling variable $\tau \equiv \epsilon t^{2.5}$ in different time regions.

 In the weak and intermediate perturbation regimes discussed in the previous sections,
 the dependence of fidelity on $\epsilon $ and $t$ does not take the form of the single
 variable $\tau$.
 This is not strange, because the classical limit is achieved in the limit
 $\hbar \to 0$, which implies $\sigma \to \infty$ for whatever small but fixed $\epsilon $.
 Therefore, it is the strong perturbation regime in which
 the decaying behavior of fidelity may have some relevance to the classical fidelity.
 Numerical results presented below indeed support this expectation.

 \begin{figure}
 \includegraphics[width=\columnwidth]{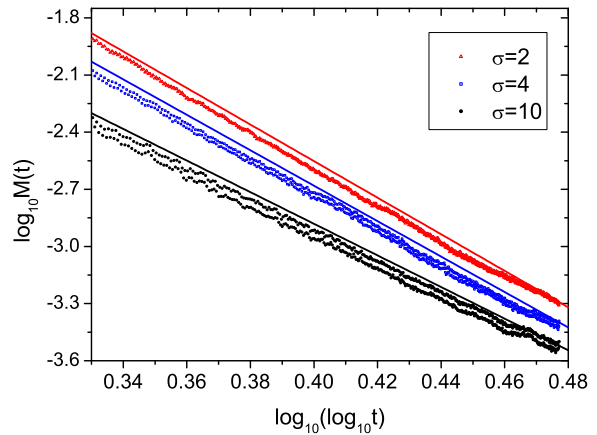}
 \caption{ (color online).
 The same as in Fig.~\ref{fig-mt-sgt1-tavg},
 with a different scale for the horizontal axis and
 for the time interval $140 < t < 1000$.
For $\sigma =4$ and 10, $\log_{10}\ov M(t)$ form two lines for each $\sigma$.
The three solid lines represent $\log_{10} M_3(t)$ given by Eq.~(\ref{loglogt}),
with $b=9.6,9.3$, and 8.3 from top to bottom.
 } \label{fig-mt-sgt1-tavg-140-1k}
 \end{figure}

 Figure \ref{fig-lglgm} shows variation of the averaged fidelity with $\log_{10}\epsilon t^{2.5}$,
 with average taken over 1000 initial Gaussian wavepackets chosen randomly.
 The initial decay of the fidelity of $\sigma =2$ and 4
 are quite close to the classical prediction $\exp (-c \epsilon t^{2.5})$.
 For longer times, the fidelity of $\sigma $ from 2 to 10 (with $\hbar $ fixed)
 is approximately a function of $\tau$, the scaling variable predicted in the classical case,
 but, the decaying behavior of fidelity
 is not the same as that of the classical fidelity, i.e., not an exponential decay.
 We found that the dependence of $\ov M(t)$ on $\hbar$ does not take the form of $\tau / \hbar$,
 i.e., $\ov M(t)$ is not a function of the single variable $(\tau / \hbar )$.

 For long times, the fidelity has large fluctuations even after averaging over 1000
 initial Gaussian wavepackets.
 The fluctuations can be much suppressed, when a further average is taken for time $t$ .
 Specifically, for each time $t$, we take average over $\ov M(t')$ for $t'$ from $t-2$ to $t+2$.
 The results are given in Fig.~\ref{fig-mt-sgt1-tavg},
 which shows that the long time decay of fidelity is slower than power law decay.
 To study the decaying behavior of the slower-than-power-law decay,
 we compare it with the function
 \be   M_3(t) = a(\log_{10} t)^{-b},  \label{loglogt} \ee
 with $a$ and $b$ as fitting parameters.
 In the time interval $140 < t < 1000$, the averaged fidelity can be fitted by this function,
 as shown in Fig.~\ref{fig-mt-sgt1-tavg-140-1k}, where we plot $\log_{10} M(t)$ versus
 $ \log_{10} (\log_{10} t)$.
 Further research work is needed to find analytical explanations for this slower-than-power-law decay
 of fidelity.

\vspace{1cm}

 \section{Conclusions and Discussions}

 We present numerical results on fidelity decay in the triangle map with linear instability.
 Three regimes of fidelity decay has been found with respect to the perturbation strength:
 weak, intermediate and strong.
 At weak perturbation, the fidelity decays like $\exp (-c \sigma^2 t^{1.7})$.
 In the intermediate regime, the fidelity has an exponential decay
 which is approximately $\exp (-c' \sigma t)$.
 In the regime of strong perturbation, the fidelity is approximately a function of
 $\epsilon t^{2.5}$
 and decays slower than power law decay for long times.

 These results show that the fidelity in the triangle map obeys decaying laws which are
 different from those in systems with strong chaos or with regular motion.
 The difference is closely related to the weak-chaos feature of the classical triangle map.
 In which way and to what extent does weak chaos influence the fidelity decay?
 This is still an open question.
 Indeed, common features of fidelity decay in systems with weak chaos, as well as
 their explanations, should be an interesting topic for future research work.
 In particular, one may note that stretch exponential decay of fidelity has also been observed
 for wave packets which initially reside in the border between chaotic
 and regular regions in mixed-type systems \cite{WLT02}.

ACKNOWLEDGMENTS. The author is very grateful to G.~Casati and T.~Prosen
for valuable discussions and suggestions.
This work is partially supported by Natural Science Foundation of China Grant
No.~10775123 and the start-up funding of USTC.


\begin{thebibliography}{99}

 \bibitem{Peres84} A.~Peres, Phys.~Rev.~A {\bf 30}, 1610 (1984).
 \bibitem{nc-book} M.A.~Nielsen and I.L.~Chuang, {\it Quantum Computation and Quantum
 Information} (Cambridge University Press, Cambridge, 2000).

 \bibitem{JP01} R.A.~Jalabert and H.M.~Pastawski, Phys.~Rev.~Lett.~{\bf 86},
     2490 (2001);
\bibitem{JSB01} Ph.~Jacquod, P.G.~Silvestrov, and C.W.J.~Beenakker,
 Phys.~Rev.~E {\bf 64}, 055203(R) (2001).
 \bibitem{CLMPV02} F.M.~Cucchietti, C.H.~Lewenkopf, E.R.~Mucciolo,
    H.M.~Pastawski, and R.O.~Vallejos, Phys.~Rev.~E {\bf 65}, 046209 (2002).
 \bibitem{JAB02} Ph.~Jacquod, I.~Adagideli, and C.W.J.~Beenakker,
                 Phys.~Rev.~Lett.~{\bf 89}, 154103 (2002).
 \bibitem{BC02} G.~Benenti and G.~Casati, Phys. Rev. E {\bf 65}, 066205(2002);
 \bibitem{CT02} N.~R.~Cerruti and S.~Tomsovic, Phys.~Rev.~Lett. {\bf 88}, 054103 (2002);
 J.~Phys.~A {\bf 36}, 3451 (2003).
 \bibitem{PZ02} T.~Prosen and M.~\v{Z}nidari\v{c}, J.~Phys.~A {\bf 35}, 1455 (2002).
 \bibitem{WL02} W.~Wang and B.~Li, Phys.~Rev.~E \textbf{66}, 056208 (2002);
 \bibitem{VH03} J.~Van\'{\i}\v{c}ek and E.J.~Heller, Phys.~Rev.~E {\bf 68}, 056208 (2003).
\bibitem{STB03} P.G.~Silvestrov, J.~Tworzyd{\l}o, and C.W.J.~Beenakker,
   Phys.~Rev.~E {\bf 67}, 025204(R) (2003).
 \bibitem{WCL04} W.Wang, G.Casati, and B.Li, Phys.~Rev.~E \textbf{69}, 025201(R)(2004).
 \bibitem{Vanicek04} J.~Van\'{\i}\v{c}ek, Phys.~Rev.~E {\bf 70}, 055201(R) (2004);
  {\bf 73}, 046204 (2006); e-print quant-ph/0410205.
  \bibitem{WL05} Wen-ge Wang and Baowen Li, Phys.~Rev.~E {\bf 71}, 066203 (2005).
 \bibitem{WCLP05} Wen-ge Wang, G.~Casati, B.~Li, and T.~Prosen, Phys.~Rev.~E {\bf 71}, 037202 (2005).
 \bibitem{GPSZ06} T.~Gorin, T.~Prosen, T.H.~Seligman, and M.~\v{Z}nidari\v{c},
 Phys.~Rep.~{\bf 435}, 33 (2006) (quant-ph/0607050).



 \bibitem{PZ03} T.~Prosen and M.~\v{Z}nidari\v{c}, New J.~Phys.~{\bf 5}, 109 (2003).
 \bibitem{JAB03} Ph.~Jacquod, I.~Adagideli, and C.W.J.~Beenakker,
                 Europhys.~Lett.~{\bf 61}, 729 (2003).
 \bibitem{SL03} R.~Sankaranarayanan and A.~Lakshminarayan, Phys.~Rev.~E {\bf 68}, 036216, 2003.
 \bibitem{WH05} Y.S.~Weinstein and C.S.~Hellberg, Phys.~Rev.~E {\bf 71}, 016209 (2005).
 \bibitem{Comb05} M.~Combescure, J.~Phys.~A {\bf 38}, 2635, 2005; M.~Combescure,
 J.~Mat.~Phys.~{\bf 47}, 032102, 2006; M.~Combescure and D.~Robert, quant-ph/0510151.
 \bibitem{HBSSR05} F.~Haug, M.~Bienert, W.~P.~Schleich, T.~H.~Seligman,
 and M.~G.~Raizen, Phys.~Rev.~A {\bf 71}, 043803, 2005.
 \bibitem{WB06} S.~Wimberger and A.~Buchleitner, J.~Phys.~B {\bf 39}, L145, 2006.
 \bibitem{pre07} Wen-ge Wang, G.~Casati, and Baowen Li, Phys.~Rev.~E {\bf 75}, 016201 (2007).

\bibitem{triangle} G.~Casati and T.~Prosen, Phys.~Rev.~Lett.~{\bf 83}, 4729 (1999);
{\it ibid}.~{\bf 85}, 4261 (2000).
\bibitem{c-fid-tri} G.~Casati, T.~Prosen, J.~Lan, and B.~Li, Phys.~Rev.~Lett.~{\bf 94},
114101 (2005).

 \bibitem{HB80-q-tori} J.H.~Hannay and M.V.~Berry, Physica D {\bf 1}, 267 (1980).
 \bibitem{FMR91} J.~Ford, G.~Mantica, and G.H.~Ristow,
 Physica D {\bf 50}, 493 (1991).
 \bibitem{WB94} J.~Wilkie and P.~Brumer, Phys.~Rev.~E {\bf 49}, 1968 (1994).

 \bibitem{Haake} F.~Haake, {\it Quantum Signatures of Chaos}, 2nd ed.
 (Springer-Verlag, Berlin, 2001).

 \bibitem{EKW05} M.~D.~Esposti, S.~O'Keefe, and B.~Winn, Nonlinearity {\bf 18}, 1073 (2005).

  \bibitem{foot1}  A similar behavior of fidelity decay in a billiard is discussed in
  D.A.~Wisniacki, E.G.~Vergini, H.M.~Pastawski, and F.M.~Cucchietti,
  Phys.~Rev.~E {\bf 65}, 055206(R) (2002).

 \bibitem{WLT02} Y.S.~Weinstein, S.~Lloyd, and C.~Tsallis, Phys.~Rev.~Lett.~{\bf 89}, 214101 (2002).



 \end{thebibliography}
 \end{document}